\documentstyle[aps,prl,multicol,epsf]{revtex}
\begin{document}
\title{Percolation in random environment}

\author{R\'obert Juh\'asz$^{1,2}$ and Ferenc Igl\'oi$^{2,1,3}$}

\address{
$^1$ Institute of Theoretical Physics,
Szeged University, H-6720 Szeged, Hungary\\
$^2$ Research Institute for Solid State Physics and Optics, 
H-1525 Budapest, P.O.Box 49, Hungary\\
$^3$Laboratoire de Physique des Mat\'eriaux, Universit\'e Henri Poincar\'e (Nancy 1),
F-54506 Vand\oe uvre l\`es Nancy, France
}

\date{\today}

\maketitle

\begin{abstract}
We consider bond percolation on the square lattice with perfectly correlated
random probabilities. According to scaling considerations, mapping to a random
walk problem and the results of Monte Carlo simulations the critical behavior
of the system with varying degree of disorder is governed by new, random fixed points
with anisotropic scaling properties. For weaker disorder both the magnetization and
the anisotropy exponents are non-universal, whereas for strong enough disorder the system
scales into an {\it infinite randomness fixed point} in which the critical
exponents are exactly known.
\end{abstract}


\newcommand{\bc}{\begin{center}}
\newcommand{\ec}{\end{center}}
\newcommand{\be}{\begin{equation}}
\newcommand{\ee}{\end{equation}}
\newcommand{\beqn}{\begin{eqnarray}}
\newcommand{\eeqn}{\end{eqnarray}}

\begin{multicols}{2}
\narrowtext

\section{Introduction}

Percolation is a paradigm for random processes\cite{staufferaharony},
in which the $i$-th bond (or site) of a
regular lattice is occupied with a probability, $p_i$, which is generally taken
independent of its position, $p_i=p$. In percolation theory one is interested in
the properties of clusters, in particular in the vicinity of the percolation transition
point $p=p_c$, when clusters with diverging size are formed. Using a close analogy
with thermal phase transitions, which is based on the $Q \to 1$ limit of the ferromagnetic
$Q$-state Potts model\cite{kasteleyn}, a scaling theory has been developed and in two dimensions
many, conjecturedly exact results have been obtained by conformal field theory\cite{cardy96}
and by Coulomb-gas methods\cite{cardyziff}.

In real systems, however, the
occupation probabilities are generally inhomogeneous,
i.e., position, direction  or neighborhood dependent, and there are some correlations
between them. The effect of quenched disorder, i.e. when the occupation probabilities
are position dependent random variables, can be studied by scaling considerations.
According to the Harris criterion\cite{harris} the relevance or irrelevance of the effect
of quenched disorder on the percolation transition depends on the sign of the specific
heat exponent, $\alpha$, of the corresponding pure Potts model in the $Q \to 1$ limit.
Since in any dimension
$\alpha<0$\cite{staufferaharony}, the critical
properties of ordinary and ``random'' percolation are equivalent.
Another form of perturbations, e.g. long-range correlations between occupation
probabilities\cite{weinrieb} or anisotropy, such as in directed percolation\cite{kinzel}, however,
leads to modified critical properties.

In the present paper we consider the combined effect of
disorder, anisotropy and correlations, when the occupation probabilities
are random variables, which are perfectly correlated in a $d_d$ dimensional subspace.
This type of behavior could be relevant to describe the properties of oil or gas inside
porous rocks in oil reservoirs, when the rock has a layered structure\cite{staufferaharony}.

Models with perfectly correlated disorder play an important r\^ ole in
statistical physics and in the theory of (quantum) phase transitions. Among the early work
we mention the partially exact solution of the
McCoy-Wu model\cite{mccoywu} (which is the two-dimensional Ising model with layered
randomness) and the field-theoretical investigations by Boyanovski and
Cardy\cite{boyanovskicardy}. As a matter of fact in random quantum systems disorder is
perfectly correlated along the (imaginary) time direction, i.e. here $d_d=1$. For these
systems, in particular for random quantum spin chains, i.e. in $(1+1)$ dimension, many
new, presumably exact results have been obtained recently by a strong disorder renormalization
group (SDRG) method\cite{MDH}. It was found that for strong enough initial disorder
the critical behavior of several systems is governed by a so-called infinite randomness
fixed point (IRFP)\cite{fisher99}, with unusual scaling properties. Here we mention
recent calculations on the random
transverse-field Ising model (RTIM)\cite{fisher92,igloi02}, random quantum Potts and
clock models\cite{senthil}, random
antiferromagnetic Heisenberg spin chains\cite{fisher94,S=1} and ladders\cite{mllri} and
also non-equilibrium phase transitions in the presence of quenched disorder\cite{hiv02}. In many
cases a cross-over between
weak and strong disorder regimes has been observed and a general scaling scenario has
been proposed\cite{cli01}.

In the present paper we study percolation on a square lattice with strip
random occupation probabilities. We investigate the critical behavior of the system with
varying strength of disorder by scaling considerations, by random walk mappings and by
Monte Carlo (MC) simulations.
The structure of the paper is the following. The model and the relevant physical
quantities are introduced in  Sec. II. Investigations in the weak and strong disorder
limits are given in Sec. III., MC simulations for intermediate disorder
are presented in Sec. IV. The paper is closed by a discussion in Sec. V.

\section{The model}

\begin{figure}[tbh]
\epsfxsize=7truecm
\begin{center}
\mbox{\epsfbox{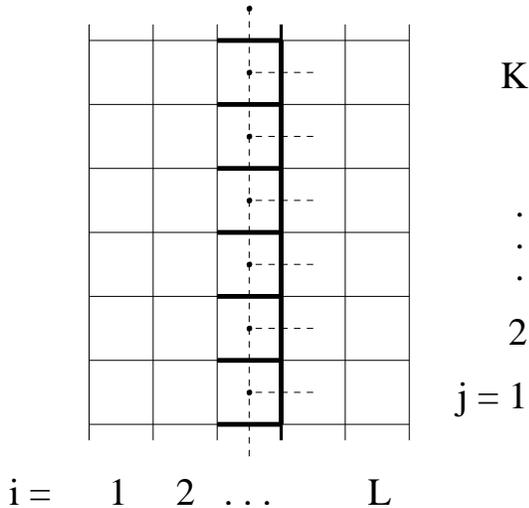}}
\end{center}
\caption{\label{fig1} Percolation on the square lattice with random bond occupation
probabilities, which are perfectly correlated along the vertical direction. The portion of
the lattice having the same occupation probability, $p$, is denoted by bold
lines, whereas the corresponding part of the dual lattice with $\tilde{p}=1-p$ is
shown by dashed lines.
} 
\end{figure}

We consider bond percolation on a square lattice with sites $\{i,j\}$,
$1 \le i \le L$ and $1 \le j \le K$, where the
occupation probabilities, $0<p<1$, are random variables, which are perfectly correlated
along vertical lines. as indicated in Fig. \ref{fig1}. If the average value of the occupation
probabilities exceeds a critical value, $\langle p \rangle>p_c$, there is a percolation
transition in the system. The value of $p_c$ can be determined by noticing that
under a duality transformation,
which maps the ordered and the disordered phases of the system into each other,
the layered structure of the system is preserved and the dual value of the local
probability is transformed as: $\tilde{p}=1-p$\cite{kinzeldomany}.
Consequently the probability distribution, $P(p)$, 
is transformed into $\tilde{P}(\tilde{p})=P(1-p)$ and the random system is self-dual, if
the probability distribution is symmetric: $P(p)=P(1-p)$ and thus the average
value of $p$ is given by $\langle p \rangle=p_c=1/2$.
Since there is one phase transition in the system, the self-duality
point corresponds to the critical point and the distance of the critical point,
$t$ is defined as:
\be
t=\langle p \rangle-p_c\;.
\label{delta}
\ee
%

In the presence of quenched disorder the mean value of a physical
observable, $\Phi$, is calculated as $[ \langle \Phi \rangle ]_{\rm av}$,
where $\langle \dots \rangle$ denotes thermal averaging for a given realization of the
disorder and $[ \dots ]_{\rm av}$ stands for disorder averaging.
In percolation the basic quantities of interest are the fractal and connectivity properties
of the largest clusters. In the following we use the concept of anisotropic
scaling\cite{binderwang},
when the correlation lengths in the two directions, (which correspond to the extensions of the
largest clusters) involve different critical exponents: $\xi_{\perp} \sim |t|^{-\nu_{\perp}}$
and $(\xi_{\parallel} \sim t^{-\nu_{\parallel}}$. Thus the anisotropy exponent
\be
z=\frac{\nu_{\parallel}}{\nu_{\perp}}\;
\label{z}
\ee
is generally different from one.
In the ordered phase, $t>0$, the number of points belonging to
the infinite cluster, $N_0$, scales around the transition point as:
\be
N_0=LK t^{\beta} \tilde{N}(L t^{\nu_{\perp}},K t^{\nu_{\parallel}})\;,
\label{N_0}
\ee
where $\beta$ is the critical exponent of the order parameter. At the critical point, $t=0$, 
fixing the ratio $K/L^z=O(1)$ we obtain:
\be
N_0 \sim L^{d_{\perp}} \sim K^{d_{\parallel}}\;,
\label{D}
\ee
where the two fractal dimensions of the infinite cluster are given by:
\be
d_{\perp}=1+z-\beta/\nu_{\perp}\;,
\label{d_f}
\ee
and $d_{\parallel}=d_{\perp}/z$.
The distribution of cluster sizes, $R(N)$, at the critical point asymptotically
behaves as:
\be
R(N) {\rm d} N= N^{-\tau} \tilde{R}(N/L^{d_{\perp}}) {\rm d} N\;,
\label{PN}
\ee
where $\tau=2+\beta/(\nu_{\perp} d_{\perp})$. This relation can be obtained by generalizing the
similar result for ordinary percolation\cite{staufferaharony}.

Correlation between two sites with coordinates, $\{i_1,j_1\}$ and $\{i_2,j_2\}$,
is defined as the expectation value of the connectivity,
$\delta(\{i_1,j_1\},\{i_2,j_2\})$, which is $1$, if the two sites belong to the
same cluster and zero otherwise. Here we mainly consider correlations in the
perpendicular direction
\be
C_{\perp}(i_1,i_2)=\frac{1}{K} \sum_{j=1}^{K}
[\langle \delta(\{i_1,j\},\{i_2,j\}) \rangle]_{\rm av} \;,
\label{Cperp}
\ee
where an average over the vertical coordinate, $j_1=j_2=j$ is also performed.
When correlations in the bulk are calculated we use periodic boundary
conditions (b.c.), (thus $i=L+1\equiv 1$), take maximal
distance between the sites, $i_2=i_1+L/2$, and
average over the position $i_1$.
The average bulk correlations, calculated in this way,
scale at the critical point as:
\be
C^b_{\perp}(L) \sim L^{-\eta_{\perp}}\;,
\label{Cb}
\ee
where $\eta_{\perp}=2 \beta/\nu_{\perp}$. We also considered the system with free
boundaries at $i=1$ and $i=L$ and calculated the correlations between two
surface sites. This end-to-end correlation function at the critical point
asymptotically behaves as:
\be
C_{\perp}(1,L) \equiv C^s_{\perp}(L) \sim L^{-\eta^s_{\perp}}\;,
\label{Cs}
\ee
where the decay exponent, $\eta^s_{\perp}$, is related to the surface fractal
properties of the infinite cluster.
Closing this section we quote the values of the critical exponents for two-dimensional
ordinary percolation\cite{staufferaharony}:
\be
\nu^{(0)}=4/3,\quad \eta^{(0)}=5/24,
\quad \eta^{s(0)}=2/3\;.
\label{exp0}
\ee

\section{Strength of disorder: Limiting cases}

The strength of disorder, $\Delta$, is related to the broadness of the probability
distribution, $P(p)$. In terms of the integrated probability distribution,
$\Pi(p)=\int_0^p P(p') {\rm d} p'$, we introduce the probabilities, $p_{1/4}$
and $p_{3/4}$ with the definitions: $\Pi(p_{1/4})=1/4$ and
$\Pi(p_{3/4})=3/4$. Since the central half of the distribution is located
in the region: $p_{1/4} \le p \le p_{3/4}$ its relative width is measured by:
\be
\Delta=\frac{p_{3/4}-p_{1/4}}{1-p_{3/4}+p_{1/4}}\;,
\label{Delta}
\ee
what we can identify with the strength of disorder.

In this paper we used two specific forms of the distribution.
For the bimodal distribution ($0 \le q \le 1/2, \overline{q}=1-q$):
\be
P_{bin}(p)=\frac{1-t}{2} \delta(p-q)+\frac{1+t}{2} \delta(p-\overline{q})\;,
\label{bimodal}
\ee
the critical point is located at $t=0$ and the strength of disorder is
given by $\Delta_{bin}=(1-2q)/2q$. Thus, as
expected the bimodal disorder is weak for $q \approx 1/2$ and strong for
$q \ll 1/2$.

The other distribution we use has a power-law form:
\be
P_{pow}(p)=\frac{1}{D 2 \overline{p}}\left( \frac{p}{\overline{p}}\right)^{-1+1/D} \qquad 0 < p <
\overline{p} < 1\;,
\label{prob} 
\ee
and $P_{pow}(1-p)=P_{pow}(p)\overline{p}/(1-\overline{p})$, for $\overline{p}<p<1$. The distance
from the critical point is measured by $t=(\overline{p}-1/2)/(D+1)$,
and for $\overline{p}=1/2$, i.e. for $t=0$ the distribution is indeed symmetric.
In this case the strength of disorder is given by: $\Delta_{pow}=2^D-1$. Thus
for $D=0$ we recover the ordinary percolation and the strength of disorder is
monotonically increasing with $D$. Therefore $D$ will be often
called as the disorder parameter of the distribution.

\subsection{Weak disorder}

In the limit of weak disorder one usually decides about the relevance-irrelevance of the
perturbation by performing a stability analysis at the ordinary percolation fixed
point. Generalizing the method by Harris\cite{harris} the
cross-over exponent due to correlated disorder is calculated as:
\be
\phi=2-\nu^{(0)}=2/3\;,
\label{harris}
\ee
where we used $\nu^{(0)}=4/3$ in Eq.(\ref{exp0}). Since $\phi>0$, even weak
correlated disorder is a relevant perturbation, thus a new random fixed point is
expected to control the critical behavior of the model.

\subsection{Strong disorder: Mapping to random walks}

\begin{figure}[tbh]
\epsfxsize=7truecm
\begin{center}
\mbox{\epsfbox{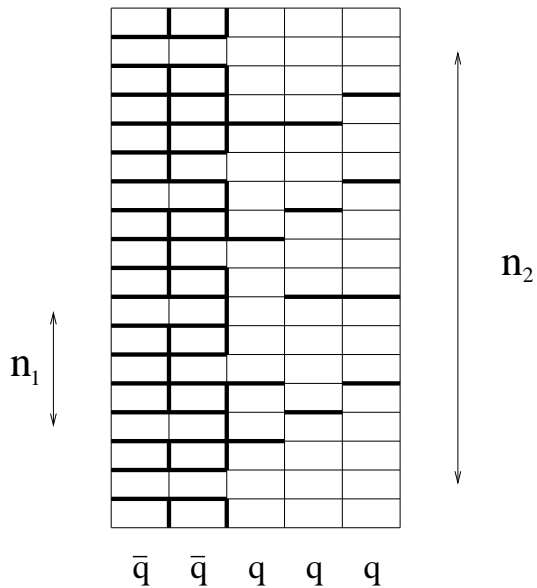}}
\end{center}
\caption{\label{fig2} Structure of the percolation cluster in the extreme bimodal
distribution with $p_1=p_2=\overline{q}$ and $p_3=p_4=p_5=q$ (here with $q \approx 1/4$).
In a layer with extremely large (small) probability there are connected (empty) units of
typical length $l \sim 1/q$. The number of sites of the connected cluster at the other
surface of a strip of width, $k$,
$n_k$ is given by: $n_1 \sim 1/q$, $n_2 \sim 1/q^2$, $n_3 \sim 1/q$ and $n_4=O(1)$ (see text).
In the limit $q \to 0$ the cluster ends at $k=4$, thus $n_5=0$.}
\end{figure}

Next we turn to study the behavior of the system for extremely strong disorder using
the bimodal distribution in Eq.(\ref{bimodal}) in the limit $q \to 0$.
In this limiting case the percolation in a given layer with a probability $p_i$
has a simple, anisotropic structure (for an illustration see Fig. \ref{fig2}).
If this probability is extremely
large, $p_i=\overline{q}$, then here almost all bonds are occupied, except of a very
small fraction of $q$. Since the typical distance between two non-occupied bonds is
$l \sim 1/q$, the cluster in the $i$-th layer is composed
of long connected units of typical size $l$. On the other hand, if the probability
is extremely small, $p_j=q$, then almost all bonds in this layer are unoccupied,
except of a very small fraction of $q$. Since the typical distance between two
occupied bonds is $l \sim 1/q$, the cluster in the $j$-th
column is composed from long empty units of typical size $l$. Notice the duality in
the structure of the two types of column.

With this prerequisite we consider the
order-parameter in the surface column, $m_s(L)$, which is the fraction of surface sites
belonging to a cluster of horizontal extent $L$. In order to make a statement about the
value of $m_s(L)$ we consider parallel strips of width $k \le L$ and introduce the
quantity, $n_k$, as the typical number of bonds at the $k$-th (i.e. surface) column of a cluster,
which is connected to the other surface of the strip. Starting with $k=1$ we have two
possibilities. For extremely small probability, $p_1=q$, there is no surface cluster
in the system, thus we have $n_1=0$. 
Otherwise, for $p_1=\overline{q}$, a surface site is connected to all sites of a
``connected unit'' of length $l$, thus we have $n_1 \sim l \sim 1/q$.
For $k=2$, if the probability is extremely large in the second layer, too, 
$p_1=p_2=\overline{q}$, then a surface cluster extends up to the second layer and
its vertical size, which is given by $n_2$, can be estimated as follows (see Fig. \ref{fig2}).
The end of a cluster is signalled by the fact that in both columns unoccupied
bonds are in neighboring positions, which happens with a probability $q^2$,
from which the typical size of a cluster $n_2 \sim 1/q^2 \sim n_1/q$, follows.
Repeating this argument for $p_i=\overline{q}$, $i=1,2,\dots,k$ we obtain
$n_k \sim 1/q^k \sim n_{k-1}/q$. Now having a small probability at the following
layer, $p_i=\overline{q}$, $i=1,2,\dots,k$ and $p_{k+1}=q$, then only a fraction of
$q$ of the sites $n_k$ have a further connection, thus $n_k$ will be reduced by a
factor $q$ giving $n_{k+1} \sim n_k q$. Inclusion of any further layer with an
extremely small probability will reduce $n_j$ by a factor of $q$, until we arrive
at $n_{j'}=O(1)$, when for the next small probability layer we have $n_{j'+1}=0$, thus
the surface cluster ends at this distance.

\begin{figure}[tbh]
\epsfxsize=7truecm
\begin{center}
\mbox{\epsfbox{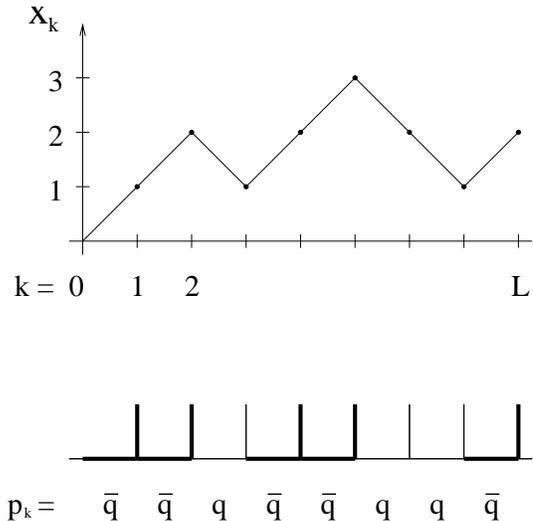}}
\end{center}
\caption{\label{fig3} Illustration of the RW mapping of percolation for a given realization of the
extreme binary distribution. Layers with high, $\overline{q}$, (low, $q$) probability are drawn
by thick (thin) lines and the corresponding RW makes a step of unit length upwards
(downwards). The position of the RW, in the $k$-th step, $X_k$, is related to, $n_k$, the number
of typical sites in the $k$-th layer of percolation, which are connected to a
given surface site as $n_k \sim q^{-X_k}$. The surface cluster extends to a distance, $L$,
if $X_k \ge 0$, for all $k=1,2, \dots L$, thus the RW has a surviving character.}
\end{figure}

From this example we can read that the $n_k$ numbers are either integer powers of
$1/q$, $n_k \sim 1/q^{X_k}$, for $X_k=0,1,\dots$, or $n_k=0$, if formally $X_k<0$. Furthermore,
we have the transformation rules:
\be
n_{k+1} \sim \left\{ \begin{array}{ll}
n_k/q, & p_{k+1}=\overline{q} \cr
n_k q, & p_{k+1}=q \cr
\end{array}
\right. 
\label{n_k}
\ee
where in the second case $n_{k+1}=0$, if $n_k=O(1)$.
At this point we can formulate the condition that the surface magnetization in a given
sample (in a rare realization) is $m_s(L)=O(1)$, if $n_k \ge O(1)$, for all
$k=1,2, \dots L$. For all other cases $m_s(L)=0$. Consequently to calculate
the {\it average value} of $m_s(L)$ it is enough to find the fraction of rare realizations,
$\rho_L^s$, for which $m_s(L)=O(1)$, since $[m_s]_{\rm av}\sim \rho_L^s$.
To calculate $\rho_L^s$ we use a random walk (RW) mapping (see an illustration in Fig. \ref{fig3}),
in which
to each disorder realization we assign a one-dimensional RW, which starts at $X_0=0$
and takes its $k$-th step upwards, $x_k=1$  (downwards, $x_k=-1$ ) if the corresponding bond
occupation probability is extremely large, $p_k=\overline{q}$ (extremely small, $p_k=q$).
The position of the walker at the
$k$-th step, $X_k=\sum_{i=1}^k x_k$ is related to $n_k$ as
$n_k \approx q^{-X_k}$. Then, as argued before, the surface cluster extends up to
a vertical distance, $L$, if $X_k \ge 0$, for every $k=1,2, \dots L$, i.e. the
RW has a surviving character. 

At the critical point of the percolation problem, $t=0$, the corresponding RW
is unbiased, and the fraction of surviving $L$-step RW-s
scales as $\rho_L^s \sim L^{-1/2}$. Now the fraction of clusters which
connect the two free boundaries of the strip over a distance $L$, and thus
contribute to the average end-to-end correlations in Eq.(\ref{Cs}),
is given by $(\rho^s_{L/2})^2$, since at each site there should be an independent
percolating surface cluster, which meet in the middle of the system.
Consequently the average end-to-end correlations at the critical point scale as
$C_{\perp}^s(L) \sim L^{-1}$ thus the corresponding decay exponent in the strong
disorder limit is given by:
\be
\eta^{s,(\infty)}_{\perp}=1\;.
\label{etas_inf}
\ee
%

Using the RW mapping one can easily estimate the perpendicular size of the
percolating clusters, which is given by
$\xi_{\parallel}(L) \sim n_{L/2} \sim q^{X_{L/2}}$. Since the transverse fluctuations
of unbiased surviving RW-s scale as $X_{L/2} \sim L^{1/2}$ we obtain in the strong
disorder limit
\be
\ln \xi_{\parallel} \sim \xi_{\perp}^{1/2} \;.
\label{z_inf}
\ee
Consequently the anisotropy exponent, $z$, in Eq.(\ref{z}) is formally infinite
for strong disorder.

Another results can be simply obtained by noticing that the same type of RW mapping
applies to the one-dimensional RTIM\cite{igloirieger98}, too,
so that we can simply borrow the results obtained in this case.

For bulk correlations one should consider the fraction of realizations, $\rho_L$,
for which a given bulk site belongs to a connected cluster of vertical size, $L$.
As was shown in\cite{riegerigloi99} for these realizations the {\it thermal average} of
the position of the RW has a surviving character. The fraction of these walks is
given by\cite{fdm,riegerigloi99}: $\rho_L \sim L^{-(3-\sqrt{5})/4}$, consequently
the critical average bulk correlations being $C^b_{\perp}(L) \sim (\rho_L)^2$
have a decay exponent
\be
\eta_{\perp}^{(\infty)}=\frac{3-\sqrt{5}}{2}\;,
\label{eta_inf}
\ee
in the strong disorder limit.

Finally, outside the critical point the mapping is related to a biased RW, with a finite
drift velocity, which is proportional to $t$. From the surviving probability of
biased RW-s one obtains for the correlation length critical exponent\cite{igloirieger98}:
\be
\nu_{\perp}^{(\infty)}=2\;,
\label{nu_inf}
\ee
The scaling exponents and relations in Eqs.(\ref{etas_inf}), (\ref{z_inf}),
(\ref{eta_inf}) and (\ref{nu_inf}) are identical with those of the IRFP of
the one-dimensional RTIM\cite{fisher92}, which is known to control the critical
behavior of several other random quantum spin chains\cite{senthil,cli01} and non-equilibrium
phase transitions in the presence of quenched disorder\cite{hiv02}.
At this point our next question is about the region of attraction of the IRFP. 
For the RTIM, where the RW
mapping can be generalized for weaker disorder, any small amount of randomness seems to
bring the system into the IRFP\cite{fisher92}, which claim is checked by intensive
numerical calculations\cite{youngrieger,igloirieger98,riegerigloi99}.
There are, however, several other models (random quantum clock-model,
Ashkin-Teller model\cite{cli01}, directed percolation\cite{hiv02}, $S=1$ random
antiferromagnetic spin chains\cite{S=1}, etc.) where
weak disorder is not sufficient to bring the system into the IRFP. 
In these cases either the pure systems fixed point stays stable against weak
disorder perturbations or the competition between (quantum)
fluctuations and weak quenched disorder leads to conventional random scaling behavior.
For the random percolation problem the latter scenario is likely to happen, since
the RW mapping can not be extended for small disorder. (The transformation law
for the connected sites, $n_k/n_{k-1} \approx q$ or $1/q$, does not hold around $q \approx 1/2$.)
We are going to study this issue numerically by MC
simulations in the next Section.

\section{Numerical results}

For intermediate strength of disorder we studied the percolation by MC simulations. 
Since the critical properties of the problem are related to the connectivity
properties of clusters for this purpose we implemented the standard Hoshen-Koopelman
labelling algorithm\cite{hk}. To decide about the shape of the lattice
one should take into account the expected anisotropic scaling properties of the
system, since the scaling functions, as in Eq.(\ref{N_0}) depend on the ratio
$r=L^z/K$, where $z$ is an unknown parameter. To overcome this difficulties
we used a strip-like geometry, when $K \gg L$, thus $r \approx 0$ for
all strip widths. In practice we had $K=10^5$, went up to $L=64$ and imposed
periodic b.c. in the vertical direction.
For the distribution of the disorder we used the power-law form in
Eq.(\ref{prob}), which has already been turned out successful in similar investigations for
random quantum spin chains\cite{cli01}. Since averaging in the vertical
direction in Eq.(\ref{Cperp}) (and also in the horizontal direction for bulk
correlations) is equivalent to a partial average over quenched disorder it was enough to
consider only a limited number ($\sim 10-20$) realizations.

\begin{figure}[tbh]
\epsfxsize=8truecm
\begin{center}
\mbox{\epsfbox{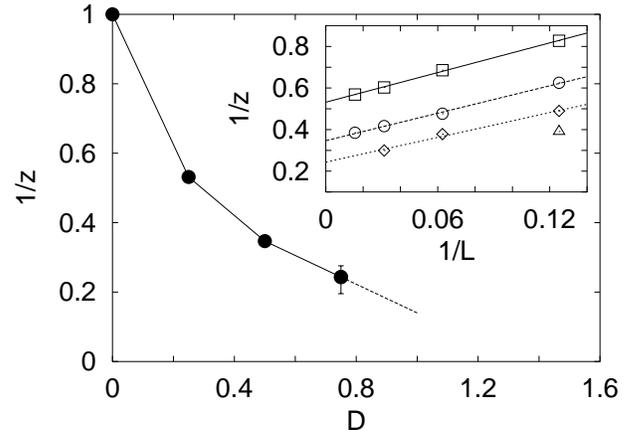}}
\end{center}
\caption{\label{fig4} Estimates for the anisotropy exponent for different
strength of disorder. The straight lines connecting the points are guide
to the eye, for $D>D_{\infty}\approx 1.2-1.5$ the anisotropy exponent
is possibly divergent. In the inset extrapolation of the size-dependent
effective anisotropy exponents is shown, for $D=0.25, 0.5, 0.75$ and $1.0$,
up to dawn.}
\end{figure}

First, we determine the anisotropy exponent, $z$, by calculating the probability distribution
of clusters in Eq.(\ref{PN}). While the decay exponent, $\tau$ in Eq.(\ref{PN})
has only a weak anisotropy dependence, the scaling function $\tilde{R}(y)$
turned out to be sensitive of the value of $z$. As we noticed in the numerical
calculations $\tilde{R}(y)$ has two different regimes. For smaller values of
the parameter, $y=N/L^z<y^*$, the finite size effects are negligible and
the scaling function is approximately constant. For $y>y^*$, when the largest
clusters touch the boundaries, the scaling function has a characteristic variation.
Measuring the position
of $y^*$ for different widths, $L$, we obtained a series of effective anisotropy
exponents, which are then extrapolated to $L \to \infty$, as shown in the inset to
Fig. \ref{fig4}. This procedure is repeated for several disorder parameters and the extrapolated
anisotropy exponents are plotted in Fig. \ref{fig4}, Unfortunately, with this method we could not
go to very strong disorder, while the cross-over region can not be clearly located
for $D>1$. However, it is clear from the available data that $z$ is monotonically
increasing with the strength of disorder and it is likely
that $z$ will be divergent for $D>D_{\infty}\approx 1.2-1.5$.

\begin{figure}[tbh]
\epsfxsize=8truecm
\begin{center}
\mbox{\epsfbox{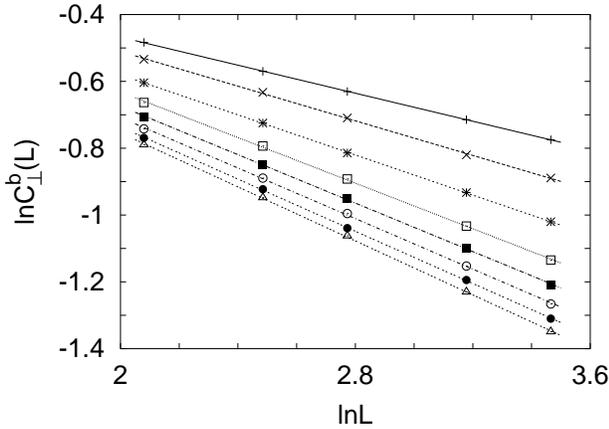}}
\end{center}
\caption{\label{fig5} Average bulk correlations vs. the width of the strip for
different strength of disorder, from $D=0$ to $D=1.75$ in units of $0.25$ from up
to dawn. The typical error is generally smaller than the size of the symbols for small $D$, 
whereas for larger $D$ it is at most twice of the size of symbols.
The straight lines are least-square fits.}
\end{figure}

In order to obtain more information about the critical behavior of the system we have
calculated the bulk and the end-to-end average correlation functions at the
critical point, as defined in Eqs.(\ref{Cb}) and (\ref{Cs}), respectively.
In Fig. \ref{fig5} the average bulk correlations, $C_{\perp}^b(L)$,
vs. $L$ is drawn in a log-log plot. The slope of the curves, which is related
to the decay exponent, $\eta_{\perp}$, has a disorder dependence.

\begin{figure}[tbh]
\epsfxsize=7truecm
\begin{center}
\mbox{\epsfbox{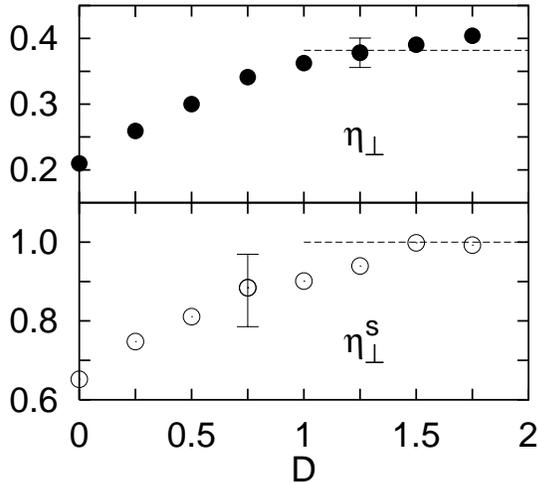}}
\end{center}
\caption{\label{fig6} Bulk ($\eta_{\perp}$) and surface ($\eta_{\perp}^s$)
decay exponents versus the strength of disorder.
Values at the IRFP, as given in Eqs.(\ref{eta_inf}) and
(\ref{etas_inf}) are denoted by dashed lines. Two typical error
bars are also indicated. }
\end{figure}

The exponents, calculated in this way together with the decay exponent of the
end-to-end correlations, $\eta_{\perp}^s$, are plotted in Fig. \ref{fig6}.
As seen in Fig. \ref{fig6} both exponents are monotonously increasing with the strength of
disorder and tend to saturate at the respective IRFP values, given in Eqs.(\ref{eta_inf})
and (\ref{etas_inf}). The value of disorder strength, where the saturation takes
place, within the error of the calculation, is the same for the two exponents and it
is compatible with the estimate, $D_{\infty}$, as
calculated from the divergence of the dynamical exponent in Fig. \ref{fig4}.

We can thus conclude that the critical behavior of the random percolation process
has a weak-to-strong disorder cross-over. For weaker disorder, $D<D_{\infty}$,
what we call the {\it intermediate disorder regime}, the critical behavior of the
system is controlled by a line of conventional fixed points. Here the anisotropy
exponent is finite, and together with the order-parameter exponents, $\eta_{\perp}$
and $\eta_{\perp}^s$, monotonously increasing with the strength of disorder.
In the {\it strong disorder regime}, $D>D_{\infty}$, the critical behavior of
the system is controlled by the IRFP. Here the anisotropy exponent is formally
infinity and the other critical exponents have no disorder dependence.

\begin{figure}[tbh]
\epsfxsize=8truecm
\begin{center}
\mbox{\epsfbox{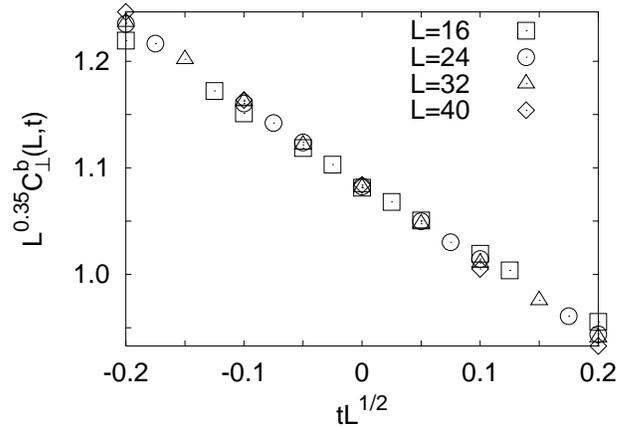}}
\end{center}
\caption{\label{fig7} Scaling plot of the bulk correlation function with
$\nu_{\perp}=2$ at a disorder strength $D=0.75$.}
\end{figure}

The non-universal nature of the critical behavior in the intermediate disorder regime is
possibly connected to the presence of a marginal operator, which should have vanishing
anomalous dimension, $x_e=0$, in the entire disorder range, $0<D<D_{\infty}$.
In our case the disorder perturbation is connected to the local energy-density
operator, for which the marginality condition, according to the Harris criterion
in Eq.(\ref{harris}) requires the condition $\phi=0$, thus
$\nu_{\perp}=2$. To verify this scenario we have calculated
the average bulk correlation function, $C_{\perp}^b(L,t)$, outside the critical
point, at a disorder strength, $D=0.75$, which is in the middle of the intermediate
disorder regime. 
According to scaling considerations
\be
C_{\perp}^b(L,t)=L^{-\eta_{\perp}} \tilde{C}(tL^{1/\nu_{\perp}})\;,
\label{Cb_scal}
\ee
thus from an optimal scaling collapse $\nu_{\perp}$ can be determined. As shown
in Fig. \ref{fig7} the scaling behavior of $C_{\perp}^b(L,t)$ is compatible with
the conjectured value of $\nu_{\perp}=2$ and thus with the marginality
condition. 

\section{Discussion}

In this paper bond percolation is studied on the square lattice with strictly
correlated, layered randomness. The phase diagram of the problem as a function
of the strength of disorder contains two regions. For strong enough disorder
the critical properties of the model are controlled by an IRFP, the
properties of which are exactly known by a RW mapping. For weaker disorder,
in the intermediate disorder regime the critical behavior is found to be
controlled by a line of conventional random fixed points, where both the
anisotropy exponent and the order-parameter exponents are disorder dependent.
The correlation length exponent, however, stays constant at its marginal value.

This type of critical behavior is very similar to that obtained in a class
of random quantum spin chains\cite{cli01,hiv02}. This close similarity
can be understood by noting the relation between percolation and the
$Q \to 1$ limit of the $Q$-state ferromagnetic Potts model.
With layered randomness the two-dimensional Potts
model in the Hamiltonian limit\cite{kogut} is equivalent to a quantum
Potts chain, with random couplings, $J_i$, and transverse fields, $h_i$.\cite{turbanigloi02},
the critical behavior of which can be studied by the SDRG method\cite{senthil}.
In this procedure the couplings and transverse fields are put in descending order
and the strongest terms are successively decimated out, whereas neighboring terms
are replaced by renormalized values. Decimating the
strongest coupling, say $J_2$, yields a new effective spin cluster in a renormalized
transverse field of strength:
\be
\tilde h=\frac{2}{Q}{h_1 h_2 \over J_2}\;,
\label{recursion}
\ee
where $h_1$ and $h_2$ are the original transverse fields acting at the two end-spins
of $J_2$.
Similarly, if, the spin in the strongest transverse field, $h_2$, is decimated out, then
a new renormalized coupling is generated between remaining spins, which is of the form in
Eq.(\ref{recursion}), by interchanging $h_i \leftrightarrow J_i$, which is due
to duality.

If the disorder is strong enough, so that the system under renormalization is
in the attractive region of the IRFP, the model specific prefactor $2/Q$
in Eq.(\ref{recursion}) does not matter and the critical properties are
universal. The region of strong attraction of the IRFP, however, is limited
by $Q=2$, i.e. for the RTIM. For smaller values of $Q$, like in percolation, when
the prefactor in Eq.(\ref{recursion}) is larger than one,
for weak disorder some renormalized couplings
and transverse fields are larger than the decimated ones. 
If this happens frequently, i.e. when the disorder is too weak, then the SDRG method
is no longer valid and the critical behavior of the model is expected to be controlled
by a conventional random fixed point. This is exactly what we obtained by MC
simulations.

We close our paper with two remarks. First, for strong enough disorder the critical
behavior of both ordinary and directed percolation\cite{hiv02} is controlled by
the same IRFP, thus the original anisotropy between the two pure problems does not
make any influence about the (strongly) random critical behavior. Our second remark
concerns possible Griffiths effects in the random percolation problem. Using the analogy
with random quantum spin chains for strong disorder some dynamical quantities of
the random percolation problem are singular also outside the critical point.
For example the susceptibility in a uniform field, $H$ diverges as $\chi \sim H^{-1+1/z'}$,
and the vertical correlation function decays algebraically as $C_{\parallel}(l) \sim l^{-1/z'}$,
where $z'$ is a finite dynamical exponent, which depends on the distance of the
critical point.

We are indebted to Jae Dong Noh, Heiko Rieger and Lo\"{\i}c Turban for stimulating
discussions. This work has been supported by the Hungarian National
Research Fund under  grant No OTKA TO34183, TO37323,
MO28418 and M36803, by the Ministry of Education under grant No FKFP 87/2001,
by the EC Centre of Excellence (No. ICA1-CT-2000-70029) and the numerical
calculations by NIIF 1030. The Laboratoire de Physique des Mat\'eriaux
is Unit\'e Mixte de Recherche No 7556.

\end{multicols}
\end{document}